\documentclass[conference]{IEEEtran}
\IEEEoverridecommandlockouts
\usepackage{cite}
\usepackage{amsmath,amssymb,amsfonts}
\usepackage{algorithmic}
\usepackage{graphicx}
\usepackage{textcomp}
\usepackage{xcolor}
\usepackage{algorithm}
\usepackage{algorithmic}
\usepackage{array}
\usepackage{booktabs}
\usepackage{subfig}
\def\BibTeX{{\rm B\kern-.05em{\sc i\kern-.025em b}\kern-.08em
    T\kern-.1667em\lower.7ex\hbox{E}\kern-.125emX}}
\begin{document}

\title{MetaBGM: Dynamic Soundtrack Transformation For Continuous Multi-Scene Experiences With Ambient Awareness And Personalization}

\author{\IEEEauthorblockN{Haoxuan Liu\IEEEauthorrefmark{3}\IEEEauthorrefmark{1}\thanks{\IEEEauthorrefmark{1}These authors contributed equally to this work.}, Zihao Wang\IEEEauthorrefmark{3}\IEEEauthorrefmark{1}, Haorong Hong\IEEEauthorrefmark{3}, Youwei Feng\IEEEauthorrefmark{3} \\ Jiaxin Yu\IEEEauthorrefmark{3}, Han Diao\IEEEauthorrefmark{3}, Yunfei Xu\IEEEauthorrefmark{2} and Kejun Zhang\IEEEauthorrefmark{3}\IEEEauthorrefmark{4}\thanks{\IEEEauthorrefmark{4}Corresponding author(s). Email(s): zhangkejun@zju.edu.cn}}

\IEEEauthorblockA{\IEEEauthorrefmark{3}College of Computer Science and Technology, Zhejiang University, Hangzhou, China\\
\IEEEauthorrefmark{2}AI Center ,OPPO Technology Co. LTD., Beijing, China\\
Email: (liuhaoxuan, carlwang, 3200102545, 22421370, yujx, 3200102306, zhangkejun)@zju.edu.cn}, xuyunfei@oppo.com
}
\maketitle

\begin{abstract}
This paper introduces MetaBGM, a groundbreaking framework for generating background music that adapts to dynamic scenes and real-time user interactions. We define multi-scene as variations in environmental contexts, such as transitions in game settings or movie scenes. To tackle the challenge of converting backend data into music description texts for audio generation models, MetaBGM employs a novel two-stage generation approach that transforms continuous scene and user state data into these texts, which are then fed into an audio generation model for real-time soundtrack creation. Experimental results demonstrate that MetaBGM effectively generates contextually relevant and dynamic background music for interactive applications.
\end{abstract}


\section{INTRODUCTION}
\label{sec:intro}
Music, as a universal cultural artifact, transcends linguistic boundaries, serving as a potent medium of expression. Traditional music composition, however, demands considerable expertise and time. Recent advancements in datasets and audio generation models, exemplified by frameworks like MusicLM\cite{agostinelli_musiclm_2023}, TANGO\cite{ghosal_text--audio_2023}, MusicGen\cite{copet_simple_2024}, WavJourney\cite{liu_wavjourney_2023}, MusicLDM\cite{chen_musicldm_2023}, and AudioLDM\cite{liu_audioldm_2024}, have revolutionized this process, enabling rapid AI-driven music composition from text prompts.

Despite these innovations, interactive music composition for dynamic multi-scene transitions remains underexplored. This is particularly crucial in automatic background music composition, where AI struggles to fluidly capture scene transitions and generate music that aligns with human expectations. Current models, which rely heavily on static music descriptions for text-to-music generation, falter in dynamic environments like film and video game scoring, where music must intricately synchronize with the narrative and atmosphere. In interactive contexts, such as gaming and interactive media, music must adapt seamlessly to real-time scene changes and user interactions, necessitating significant advancements in audio generation models to discern and respond to these shifts by producing coherent, contextually appropriate background music. Therefore, generating high-quality music descriptions becomes a critical research challenge.

Traditionally, crafting these descriptions has been the purview of skilled musicians, a process both time-consuming and resource-intensive\cite{cai_music_2020}. The advent of large language models (LLMs) presents new opportunities for generating music descriptions in complex scenarios\cite{lam_efficient_2023}. These models can transform scene and interaction data into coherent, contextually appropriate textual descriptions. Within this framework, procedural narrative\cite{griffith2018procedural} generation converts continuous scene data into descriptive texts\cite{roberts_targeting_nodate}\cite{sharma_drama_2010}, capturing dynamic shifts in scenes and character interactions\cite{riedl_interactive_2013}. For instance, SceneCraft\cite{kumaran_scenecraft_2023} showcases this by generating adaptive dialogue paths that enhance the immersive quality of digital games. Automatic music description generation then translates these narratives into corresponding music texts, providing precise inputs for audio generation models. Projects like Noise2Music\cite{huang_noise2music_2023}, MeLoDy\cite{lam_efficient_2023}, and LP\_MusicCaps\cite{doh_lp-musiccaps_2023} leverage LLMs to generate rich music descriptions, yet most research remains focused on static descriptions, overlooking challenges in multi-scene transitions and real-time interactions. Moreover, despite advances in controllable music generation, the flexibility required for real-time adaptation remains insufficient.

This study introduces a novel method to transform continuous scene data into music descriptions interpretable by audio generation models, enabling the creation of adaptive background music. We propose an innovative algorithm for interactive, multi-scene music description generation, utilizing a two-stage pipeline to produce effective music descriptions. These descriptions are then used by audio generation models to create background music that seamlessly aligns with dynamic, continuous scenes in real time. Using Minecraft as a case study, we developed a real-time backend data collection algorithm and implemented a two-stage LLM-based generation process, encompassing both procedural narrative generation and music description synthesis. Additionally, we fine-tuned open-source LLMs to optimize music description outputs.

\begin{figure*}[ht]
  \centering
  \includegraphics[width=\textwidth]{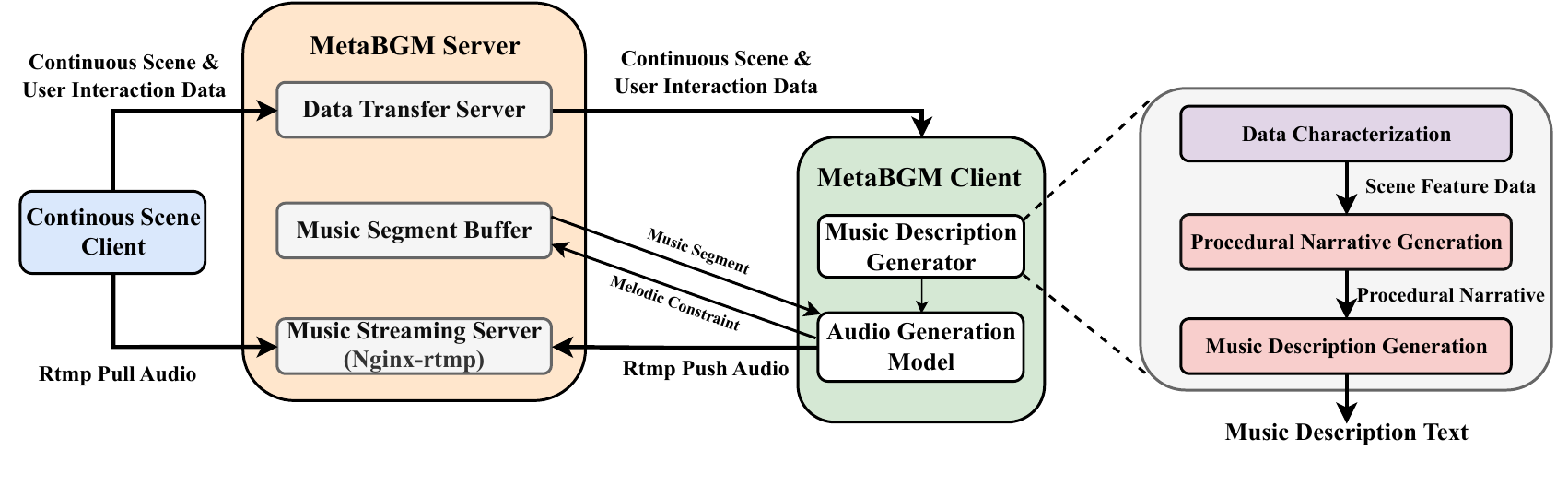}
  \vspace{-0.3cm}
  \caption{Proposed framework of MetaBGM.}
  \label{img:metabgm}
  \vspace{-0.2cm}
\end{figure*}

Our contributions introduce the task of interactive real-time background music generation and propose the MetaBGM framework, which utilizes a two-stage music description generation method. Additionally, we developed a paired dataset from dynamic scene descriptions, filling a critical research gap. MetaBGM’s experimental results demonstrate its capability to generate contextually relevant background music that adapts fluidly to continuous scene transitions while ensuring real-time precision in synchronizing with user interactions. 

\section{PROPOSED APPROACH}
\label{sec:proposed approach}
We employed the open-world continuous scene environment of Minecraft, celebrated for its intricate settings and dynamic interactions. Minecraft features over 62 distinct scenes and a plethora of player actions, offering a robust dataset for multi-scene and interaction analyses, alongside more than 50 adaptable music tracks.

\subsection{Framework of MetaBGM}
\label{subsec:metabgm framework}
As illustrated in Fig. \ref{img:metabgm}, the framework operates by capturing real-time scene and user interaction data during each time segment\cite{androulaki_hyperledger_2018}, including context and actions. This data is incorporated into the model prompt and processed by the LLM to generate procedural narrative text. This narrative is then employed to create contextually aligned music description text. The audio generation model uses this description to produce real-time background music, with each piece serving as a melodic anchor for the next segment, ensuring fluid continuity across scene transitions. The interval between segments is adjustable.

\subsection{Real-time Backend Data Acquisition}
\label{subsec:Backend Data Acquisition}
We deployed a continuous scene client-server framework using official open-source code, with the client initializing the environment locally and the server handling real-time scene and user interaction data. To support this, we engineered a real-time backend data acquisition algorithm that captures the player’s scene, status, and actions in JSON format at customizable intervals, typically every 10 seconds. The algorithm continuously aggregates data across 11 elements—biome, time, weather, temperature, health, hunger, action state, and combat state. A multi-threaded approach ensures seamless data collection, with the main thread acquiring data while auxiliary threads manage aggregation and parsing. Extracted data often includes superfluous details such as “not on fire,” “not running,” or “not sneaking,” which are only relevant in affirmative contexts and otherwise introduce unnecessary complexity. Furthermore, backend data is stored in double-precision floating-point format, imposing excessive precision that burdens the language model without enhancing insight. The significance of these data points fluctuates with context; for example, during combat, environmental data like weather is less critical, while the player’s health and actions become paramount. Conversely, in non-combat exploration, the relevance of combat data diminishes, and environmental factors gain importance. To address these variances, we developed a data characterization algorithm that prioritizes relevant information based on context, filtering out redundant details and adjusting data precision as necessary. The detailed data characterization algorithm is illustrated in Algorithm \ref{alg1}.
\begin{algorithm} 
    \renewcommand{\algorithmicrequire}{\textbf{Input:}}
    \renewcommand{\algorithmicensure}{\textbf{Output:}}
    \caption{Data Feature Processing Algorithm} 
    \label{alg1} 
    \begin{algorithmic}
        \REQUIRE{Scene Data: Data obtained from the continuous scene, formatted as a dictionary}
        \ENSURE{CharacterizedData: Processed data after feature extraction, formatted as a dictionary}
        \FOR{key, value in Scene Data.items()}
            \IF {isinstance(value, dict)}
                \FOR{inner\_key, inner\_value in value.items()}
                    \IF {`Not' in inner\_key}
                        \STATE Skip inner\_key, inner\_value pair
                    \ENDIF
                    \STATE Retain only two decimal places for inner\_value
                    \STATE CharacterizedData[key] $\gets$ new\_inner\_dict
                \ENDFOR
            \ENDIF
        \ENDFOR
        \IF {current state is combat mode}
            \STATE Retain only the information in CharacterizedData related to ["Scene", "Health", "Satiety", "Status", "Movement", "Position", "Hostile Entity", "Being Attacked"]
        \ELSE
            \STATE Retain only the information in CharacterizedData related to ["Scene", "Time", "Weather", "Temperature", "Status", "Movement", "Position"]
        \ENDIF
    \end{algorithmic} 
\end{algorithm}
\subsection{Two-Stage Music Description Generation}
\label{subsec:two-stage music description generation}
After acquiring scene and user interaction data, we adopt a two-stage method for generating music descriptions. Initially, the JSON-formatted data is converted into narrative text, which is subsequently used to create the corresponding music description. This two-stage approach, rather than direct music description generation, is employed due to the lower readability and structured nature of JSON data compared to narrative language, which can yield less varied and rich music descriptions. Furthermore, in language modeling, the Chain of Thought method indicates that incorporating intermediate reasoning steps into the model’s prompts can significantly enhance its reasoning capabilities\cite{wang_plan-and-solve_2023}. Thus, by first generating narrative text, the model attains a better comprehension of the task, leading to more accurate and contextually relevant music descriptions.

\subsubsection{Procedural Narrative Generation}
\label{subsubsection:Procedural Narrative Generation}
To enhance the readability and comprehensibility of the characterized JSON-formatted data, we drew inspiration from procedural narrative generation techniques and transformed the JSON data into natural language narrative text. This study employs a LLM to generate narrative text, capitalizing on the LLM’s robust contextual understanding and information extraction capabilities, which have achieved state-of-the-art performance in text summarization. More importantly, LLMs demonstrate creativity, enabling them to not only accurately summarize the JSON data but also generate vivid narrative descriptions, occasionally incorporating reasonable details. This methodology enriches the monotonous and rigid format of the source data, enhancing the diversity of the final music descriptions. The narrative text generation task using LLMs can generally be represented by (\ref{eq1})\cite{qiu_smile_2024}:

\begin{equation} \label{eq1}
P_{M_{\theta}}(Y \mid x) = \prod_{t=1}^{L} P_{M_{\theta}}(y_{t} \mid Y_{<t}, x)
\end{equation}

where \(M\) represents the LLM, and \(\theta\) are the model parameters. \(Y\) denotes the model's output text, which is the narrative text, and \(x\) represents the model input under specified conditions, with \(L\) being the variable output length. \(y_t\) refers to the \(t\)-th character generated by the model.

\begin{figure}[t]
    \centering
    \subfloat[Narrative text generation prompt template. The \textit{info\_str} refers to the game data converted into text.\label{fig:narrative}]{%
        \includegraphics[width=\linewidth]{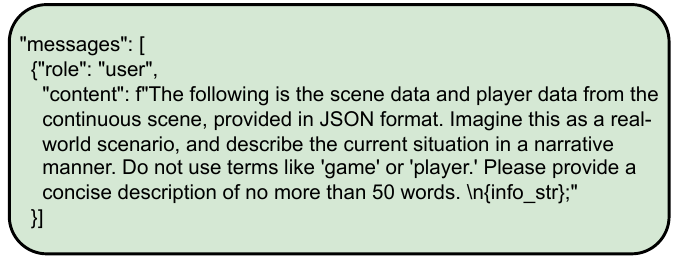}%
    }
    
    \subfloat[Music description generation prompt template. The \textit{scene} refers to the narrative text from Section \ref{subsubsection:Procedural Narrative Generation}\label{fig:musicdesc}]{%
        \includegraphics[width=\linewidth]{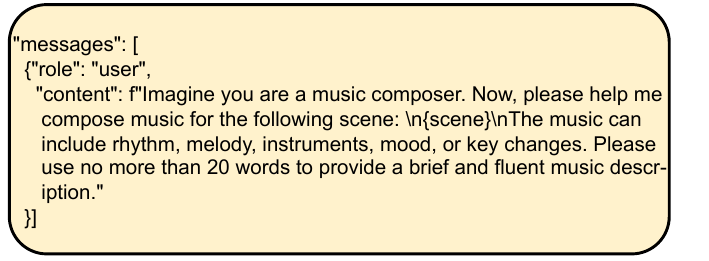}%
    }
    
    \caption{Prompt template of two-stage method for generating music descriptions.}
    \label{img:prompt}
    \vspace{-0.5cm}
\end{figure}

We employed the LLaMA-2-7B-Chat model, fine-tuned for dialogue content, to generate both narrative and music description texts. This model, with its 7 billion parameters\cite{touvron_llama_2023}\cite{touvron_llama_2023-1}, facilitates efficient deployment and adaptation. Preliminary testing confirmed its efficacy in generating narrative content. A specialized prompt template, tailored to task-specific requirements (see Fig. \ref{fig:narrative}), was devised to enhance narrative diversity. By prompting the LLM to envision the scene as a real-world scenario, the model produces vivid, imaginative descriptions, such as: “In the midday sun, amid the gravel hills swept by a gentle breeze, a healthy, well-equipped individual stands, unconcerned about hunger, moving cautiously on solid ground, away from any fire hazards.” In contrast, without this guiding prompt, the output tends to be more formulaic, e.g., “The current scene is a forest in the morning, with mild temperatures… The player’s status is dry, with full health and hunger levels…” This carefully constructed prompt enables the LLM to generate concise narrative texts, which are subsequently utilized for music description generation.

\begin{table*}[t] % [t] 选项用于将表格放在页面顶部
\centering
\caption{The results of the ablation experiment for the two-stage generation.}
\vspace{-0.2cm}
\begin{tabular}{@{}c|cccccc@{}}
\toprule
Model & B-1$\uparrow$ & B-2$\uparrow$ & B-3$\uparrow$ & B-4$\uparrow$ & METEOR$\uparrow$ & R-L$\uparrow$ \\ \midrule
LLaMA-2-7B-Chat + Direct Generation & 44.06 & 18.26 & 6.85 & 3.22 & 17.41 & 18.87 \\
LLaMA-2-7B-Chat + Two-Stage Generation & \textbf{51.66} & \textbf{23.89} & \textbf{8.78} & \textbf{3.79} & \textbf{20.86} & \textbf{20.07} \\
GPT-3.5-Turbo + Direct Generation & 50.07 & 15.62 & 4.08 & 2.01 & 22.38 & 20.1 \\
GPT-3.5-Turbo + Two-Stage Generation & \textbf{58.84} & \textbf{21.63} & \textbf{9.89} & \textbf{4.86} & 17.52 & \textbf{21.21} \\ \bottomrule
\end{tabular}
\label{tab:1}
\end{table*}

\begin{table*}
\centering
\caption{The objective evaluation scores on the six metrics across all experimental groups.}
\vspace{-0.2cm}
\begin{tabular}{@{}c|cccccc@{}}
\toprule
Model & B-1$\uparrow$ & B-2$\uparrow$ & B-3$\uparrow$ & B-4$\uparrow$ & METEOR$\uparrow$ & R-L$\uparrow$ \\ \midrule
LLaMA-2-7B & 31.05 & 12.46 & 4.46 & 2.09 & 17.97 & 16.57 \\
LLaMA-2-13B & 32.77 & 16.7 & 7.96 & 3.11 & 20.77 & 19.05 \\
ChatGLM-3-6B & 41.94 & 16.12 & 6.49 & 3.44 & 18.33 & 17.44 \\
LLaMA-2-7B-Chat + Direct Generation & 44.06 & 18.26 & 6.85 & 3.22 & 17.41 & 18.87 \\
LLaMA-2-7B-Chat + Two-Stage Generation & 51.66 & 23.89 & 8.78 & 3.79 & 20.86 & 20.07 \\
GPT-3.5-Turbo + Direct Generation & 50.07 & 15.62 & 4.08 & 2.01 & \textbf{22.38} & 20.1 \\
GPT-3.5-Turbo + Two-Stage Generation & \textbf{58.84} & 21.63 & 9.89 & 4.86 & 17.52 & \textbf{21.21} \\
MetaBGM & 54.61 & \textbf{24.13} & \textbf{10.08} & \textbf{4.9} & 15.79 & 19.36 \\
\bottomrule
\end{tabular}
\label{tab:2}
\end{table*}

\subsubsection{Music Description Generation}
\label{subsubsec:musicgeneration}
The LLaMA-2-7B-Chat\cite{touvron_llama_2023}\cite{touvron_llama_2023-1} was employed to generate music description texts, as mathematically represented in Equation (\ref{eq1}). Narrative text from the previous section was integrated into the prompt template, as shown in Fig. \ref{fig:musicdesc}, to produce the final music descriptions. The prompt design adheres to prompt engineering principles, allowing the model to assume the role of a music composer while offering potential elements for the description, thereby refining the precision of the output. Moreover, the prompts enforce a 20-word limit on the generated descriptions, a critical constraint given that these descriptions will guide audio generation models, where excessive length could hinder real-time music synthesis. For instance, a prompt designed for a forest scene yielded the following description: “The music is gentle and soothing, with a soft melody and stable rhythm. It features a blend of acoustic and electronic elements, creating a sense of relaxation and tranquility.” In summary, this section first elucidated the characterization of game data as detailed in Section (\ref{subsec:Backend Data Acquisition}), followed by the generation of narrative texts and, ultimately, the creation of music descriptions using the LLaMA-2-7B-Chat model and the customized prompt template.

\section{EXPERIMENT}
\label{sec:expcon}

\subsection{Dataset}
\label{subsec:dataset}
This study necessitates paired datasets of Scene Data-Narrative Text and Narrative Text-Music Description Text. To construct the training set, two strategies were employed. First, leveraging chatGPT-3.5-Turbo, 433 feature-rich scenarios were synthesized by varying parameters such as scene, time, weather, and player status, yielding 433 Scene Data-Narrative Text and Narrative Text-Music Description Text pairs\cite{taori2023alpaca}. Second, by employing reverse generation, we curated music descriptions for 2,000 tracks from the Song Describer Dataset\cite{manco_song_2023} and MusicCaps\cite{agostinelli_musiclm_2023}, subsequently matching them with suitable Minecraft scenes via chatGPT-3.5-Turbo, thereby generating an additional 1,539 Narrative Text-Music Description Text pairs.

\subsection{Training and Inference Parameters}
\label{subsec:exp-gr}
To enhance narrative and music description generation while preserving the core capabilities of LLMs, the LoRA\cite{hu_lora_2021} method was employed for its computational efficiency and effectiveness. The model was fine-tuned using 1,972 pairs of Scene Data - Narrative Text and Narrative Text - Music Description Text from the Section \ref{subsec:dataset}. These datasets were structured into an Instruction-Input-Output format for LoRA fine-tuning, optimizing performance during supervised learning.

\subsection{Test Set and Evaluation Metrics}
\label{subsec;testset}
To evaluate the model's efficacy, this study analyzed 13 original background music tracks from Minecraft, each tied to a specific scene. Using the LP\_MusicCaps\cite{doh_lp-musiccaps_2023} model, these tracks were transformed into 304 descriptive segments at 10-second intervals, serving as a test set due to their scene-specific nature. For assessment, we utilized established bilingual translation metrics, including BLEU-1, BLEU-2, BLEU-3, and BLEU-4, alongside the METEOR metric for translation accuracy and recall, and ROUGE-L for comparing summary and reference similarity.

\subsection{Results and Analysis}
\label{subsec:typestyle}
The baseline models encompass prominent LLMs such as Meta’s LLaMA-2-7B, LLaMA-2-7B-Chat (unfine-tuned), LLaMA-2-13B\cite{touvron_llama_2023}\cite{touvron_llama_2023-1}, Tsinghua University’s ChatGLM-3-6B\cite{du_glm_2022}, and OpenAI’s GPT-3.5-Turbo. The experiments initially compared direct music description generation with the proposed two-stage generation method, as shown in TABLE \ref{tab:1}. The two-stage approach, which first transforms JSON-formatted game data into scene descriptions before generating music descriptions, consistently outperformed direct generation across all metrics. This validates the hypothesis that incremental generation enhances logical reasoning and accuracy in large models, paralleling chain-of-thought reasoning in AI.

TABLE \ref{tab:2} encapsulates the results for all baseline models. The MetaBGM method, employing a fine-tuned LLaMA-2-7B-Chat and the two-stage generation process, exhibited substantial improvements over baseline models, with the exception of GPT-3.5-Turbo. While GPT-3.5-Turbo combined with two-stage generation marginally surpassed our method, it is crucial to note that GPT-3.5-Turbo contains over 175B parameters—25 times more than LLaMA-2-7B-Chat. Despite this discrepancy in parameter count, the fine-tuned LLaMA-2-7B-Chat demonstrated superior performance, particularly in managing conversational inputs, validating its selection for this study.

\section{CONCLUSION}
\label{conclusion}
MetaBGM is an LLM-based framework that generates background music corresponding to continuous scenes and user interaction events through a two-stage music description generation method. The key idea involves using multi-threading to obtain contextual data from continuous scenes and user interactions, calculating features, and then converting them into music description text interpretable by the audio generation model. The evaluation results demonstrate the effectiveness of this background music generation method, with objective metrics indicating that MetaBGM can produce background music that aligns with scene transitions and user events at a relatively low cost. 

\bibliographystyle{IEEEtran} % 使用IEEEtran的文献样式
\bibliography{main} % 

\begin{thebibliography}{10}
\providecommand{\url}[1]{#1}
\csname url@samestyle\endcsname
\providecommand{\newblock}{\relax}
\providecommand{\bibinfo}[2]{#2}
\providecommand{\BIBentrySTDinterwordspacing}{\spaceskip=0pt\relax}
\providecommand{\BIBentryALTinterwordstretchfactor}{4}
\providecommand{\BIBentryALTinterwordspacing}{\spaceskip=\fontdimen2\font plus
\BIBentryALTinterwordstretchfactor\fontdimen3\font minus \fontdimen4\font\relax}
\providecommand{\BIBforeignlanguage}[2]{{%
\expandafter\ifx\csname l@#1\endcsname\relax
\typeout{** WARNING: IEEEtran.bst: No hyphenation pattern has been}%
\typeout{** loaded for the language `#1'. Using the pattern for}%
\typeout{** the default language instead.}%
\else
\language=\csname l@#1\endcsname
\fi
#2}}
\providecommand{\BIBdecl}{\relax}
\BIBdecl

\bibitem{agostinelli_musiclm_2023}
A.~Agostinelli, T.~I. Denk, Z.~Borsos, J.~Engel, M.~Verzetti, A.~Caillon, Q.~Huang, A.~Jansen, A.~Roberts, M.~Tagliasacchi, M.~Sharifi, N.~Zeghidour, and C.~Frank, ``{MusicLM}: {Generating} {Music} {From} {Text},'' \emph{arXiv preprint arXiv:2301.11325}, 2023.

\bibitem{ghosal_text--audio_2023}
D.~Ghosal, N.~Majumder, A.~Mehrish, and S.~Poria, ``Text-to-{Audio} {Generation} using {Instruction}-{Tuned} {LLM} and {Latent} {Diffusion} {Model},'' \emph{arxiv preprint arXiv:2304.13731}, 2023.

\bibitem{copet_simple_2024}
J.~Copet, F.~Kreuk, I.~Gat, T.~Remez, D.~Kant, G.~Synnaeve, Y.~Adi, and A.~Défossez, ``Simple and {Controllable} {Music} {Generation},'' \emph{arxiv preprint arXiv:2306.05284}, 2024.

\bibitem{liu_wavjourney_2023}
X.~Liu, Z.~Zhu, H.~Liu, Y.~Yuan, M.~Cui, Q.~Huang, J.~Liang, Y.~Cao, Q.~Kong, M.~D. Plumbley, and W.~Wang, ``{WavJourney}: {Compositional} {Audio} {Creation} with {Large} {Language} {Models},'' \emph{arxiv preprint arXiv:2307.14335}, 2023.

\bibitem{chen_musicldm_2023}
K.~Chen, Y.~Wu, H.~Liu, M.~Nezhurina, T.~Berg-Kirkpatrick, and S.~Dubnov, ``{MusicLDM}: {Enhancing} {Novelty} in {Text}-to-{Music} {Generation} {Using} {Beat}-{Synchronous} {Mixup} {Strategies},'' \emph{arxiv preprint arXiv:2308.01546}, 2023.

\bibitem{liu_audioldm_2024}
H.~Liu, Y.~Yuan, X.~Liu, X.~Mei, Q.~Kong, Q.~Tian, Y.~Wang, W.~Wang, Y.~Wang, and M.~D. Plumbley, ``{AudioLDM} 2: {Learning} {Holistic} {Audio} {Generation} with {Self}-supervised {Pretraining},'' \emph{arXiv preprint arXiv:2308.05734}, 2024.

\bibitem{cai_music_2020}
T.~Cai, M.~I. Mandel, and D.~He, ``Music autotagging as captioning,'' in \emph{Proceedings of the 1st {Workshop} on {NLP} for {Music} and {Audio}}, S.~Oramas, L.~Espinosa-Anke, E.~Epure, R.~Jones, M.~Sordo, M.~Quadrana, and K.~Watanabe, Eds.\hskip 1em plus 0.5em minus 0.4em\relax Association for Computational Linguistics, 2020, pp. 67--72.

\bibitem{lam_efficient_2023}
M.~W.~Y. Lam, Q.~Tian, T.-C. Li, Z.~Yin, S.~Feng, M.~Tu, Y.~Ji, R.~Xia, M.~Ma, X.~Song, J.~Chen, Y.~Wang, and Y.~Wang, ``Efficient {Neural} {Music} {Generation},'' \emph{arxiv preprint arXiv:2305.15719}, 2023.

\bibitem{griffith2018procedural}
I.~Griffith, ``Procedural narrative generation through emotionally interesting non-player characters,'' Master's thesis, Linnaeus University, 2018.

\bibitem{roberts_targeting_nodate}
D.~L. Roberts, M.~J. Nelson, J.~Charles L.~Isbell, M.~Mateas, and M.~L. Littman, ``Targeting {Specific} {Distributions} of {Trajectories} in {MDPs},'' in \emph{Proceedings of the 21st National Conference on Artificial Intelligence (AAAI 06)}, 2006.

\bibitem{sharma_drama_2010}
M.~Sharma, S.~Ontañón, M.~Mehta, and A.~Ram, ``Drama {Management} and {Player} {Modeling} for {Interactive} {Fiction} {Games},'' in \emph{Computational Intelligence}, vol.~26, 2010, pp. 183--211.

\bibitem{riedl_interactive_2013}
M.~O. Riedl and V.~Bulitko, ``Interactive {Narrative}: {An} {Intelligent} {Systems} {Approach},'' in \emph{AI Magazine}, vol.~34, 2013, pp. 67--77.

\bibitem{kumaran_scenecraft_2023}
V.~Kumaran, J.~Rowe, B.~Mott, and J.~Lester, ``{SceneCraft}: {Automating} {Interactive} {Narrative} {Scene} {Generation} in {Digital} {Games} with {Large} {Language} {Models},'' in \emph{Proceedings of the AAAI Conference on Artificial Intelligence and Interactive Digital Entertainment}, 2023, pp. 86--96.

\bibitem{huang_noise2music_2023}
Q.~Huang, D.~S. Park, T.~Wang, T.~I. Denk, A.~Ly, N.~Chen, Z.~Zhang, Z.~Zhang, J.~Yu, C.~Frank, J.~Engel, Q.~V. Le, W.~Chan, Z.~Chen, and W.~Han, ``{Noise2Music}: {Text}-conditioned {Music} {Generation} with {Diffusion} {Models},'' 2023.

\bibitem{doh_lp-musiccaps_2023}
S.~Doh, K.~Choi, J.~Lee, and J.~Nam, ``{LP}-{MusicCaps}: {LLM}-{Based} {Pseudo} {Music} {Captioning},'' \emph{arxiv preprint arXiv:2307.16372}, 2023.

\bibitem{androulaki_hyperledger_2018}
E.~Androulaki, A.~Barger, V.~Bortnikov, C.~Cachin, K.~Christidis, A.~De~Caro, D.~Enyeart, C.~Ferris, G.~Laventman, Y.~Manevich, S.~Muralidharan, C.~Murthy, B.~Nguyen, M.~Sethi, G.~Singh, K.~Smith, A.~Sorniotti, C.~Stathakopoulou, M.~Vukolić, S.~W. Cocco, and J.~Yellick, ``Hyperledger fabric: a distributed operating system for permissioned blockchains,'' in \emph{Proceedings of the {Thirteenth} {EuroSys} {Conference}}, ser. {EuroSys} '18.\hskip 1em plus 0.5em minus 0.4em\relax Association for Computing Machinery, 2018, pp. 1--15.

\bibitem{wang_plan-and-solve_2023}
L.~Wang, W.~Xu, Y.~Lan, Z.~Hu, Y.~Lan, R.~K.-W. Lee, and E.-P. Lim, ``Plan-and-{Solve} {Prompting}: {Improving} {Zero}-{Shot} {Chain}-of-{Thought} {Reasoning} by {Large} {Language} {Models},'' \emph{arxiv preprint arXiv:2305.04091}, 2023.

\bibitem{qiu_smile_2024}
H.~Qiu, H.~He, S.~Zhang, A.~Li, and Z.~Lan, ``{SMILE}: {Single}-turn to {Multi}-turn {Inclusive} {Language} {Expansion} via {ChatGPT} for {Mental} {Health} {Support},'' \emph{arxiv preprint arXiv:2305.00450}, 2024.

\bibitem{touvron_llama_2023}
H.~Touvron, T.~Lavril, G.~Izacard, X.~Martinet, M.-A. Lachaux, T.~Lacroix, B.~Rozière, N.~Goyal, E.~Hambro, F.~Azhar, A.~Rodriguez, A.~Joulin, E.~Grave, and G.~Lample, ``{LLaMA}: {Open} and {Efficient} {Foundation} {Language} {Models},'' \emph{arxiv preprint arXiv:2302.13971}, 2023.

\bibitem{touvron_llama_2023-1}
H.~Touvron, L.~Martin, K.~Stone, P.~Albert, A.~Almahairi, Y.~Babaei, N.~Bashlykov, S.~Batra, P.~Bhargava, S.~Bhosale, D.~Bikel, L.~Blecher, C.~C. Ferrer, M.~Chen, G.~Cucurull, D.~Esiobu, J.~Fernandes, J.~Fu, W.~Fu, B.~Fuller, C.~Gao, V.~Goswami, N.~Goyal, A.~Hartshorn, S.~Hosseini, R.~Hou, H.~Inan, M.~Kardas, V.~Kerkez, M.~Khabsa, I.~Kloumann, A.~Korenev, P.~S. Koura, M.-A. Lachaux, T.~Lavril, J.~Lee, D.~Liskovich, Y.~Lu, Y.~Mao, X.~Martinet, T.~Mihaylov, P.~Mishra, I.~Molybog, Y.~Nie, A.~Poulton, J.~Reizenstein, R.~Rungta, K.~Saladi, A.~Schelten, R.~Silva, E.~M. Smith, R.~Subramanian, X.~E. Tan, B.~Tang, R.~Taylor, A.~Williams, J.~X. Kuan, P.~Xu, Z.~Yan, I.~Zarov, Y.~Zhang, A.~Fan, M.~Kambadur, S.~Narang, A.~Rodriguez, R.~Stojnic, S.~Edunov, and T.~Scialom, ``Llama 2: {Open} {Foundation} and {Fine}-{Tuned} {Chat} {Models},'' \emph{arxiv preprint arXiv:2307.09288}, 2023.

\bibitem{taori2023alpaca}
R.~Taori, I.~Gulrajani, T.~Zhang, Y.~Dubois, X.~Li, C.~Guestrin, P.~Liang, and T.~B. Hashimoto, ``Stanford alpaca: An instruction-following llama model,'' 2023.

\bibitem{manco_song_2023}
I.~Manco, B.~Weck, S.~Doh, M.~Won, Y.~Zhang, D.~Bogdanov, Y.~Wu, K.~Chen, P.~Tovstogan, E.~Benetos, E.~Quinton, G.~Fazekas, and J.~Nam, ``The {Song} {Describer} {Dataset}: a {Corpus} of {Audio} {Captions} for {Music}-and-{Language} {Evaluation},'' \emph{arxiv preprint arXiv:2311.10057}, 2023.

\bibitem{hu_lora_2021}
E.~J. Hu, Y.~Shen, P.~Wallis, Z.~Allen-Zhu, Y.~Li, S.~Wang, L.~Wang, and W.~Chen, ``{LoRA}: {Low}-{Rank} {Adaptation} of {Large} {Language} {Models},'' \emph{arxiv preprint arXiv:2106.09685}, 2021.

\bibitem{du_glm_2022}
Z.~Du, Y.~Qian, X.~Liu, M.~Ding, J.~Qiu, Z.~Yang, and J.~Tang, ``{GLM}: {General} {Language} {Model} {Pretraining} with {Autoregressive} {Blank} {Infilling},'' \emph{arxiv preprint arXiv:2103.10360}, 2022.

\end{thebibliography}
\end{document}